# THE SMALL SCALE
# INTEGRATED SACHS-WOLFE EFFECT [†]

Wayne Hu[1] and Naoshi Sugiyama[1,2]

[1]*Departments of Astronomy and Physics*
*University of California, Berkeley, California 94720*

[2] *Department of Physics, Faculty of Science*
*The University of Tokyo, Tokyo, 113, Japan*

The integrated Sachs-Wolfe effect (ISW) can be an important factor in the generation of Cosmic Microwave Background anisotropies *on all scales* especially in a reionized curvature or lambda dominated universe. We present an analytic treatment of the ISW effect, which is analogous to thick last scattering surface techniques for the Doppler effect, that compares quite well with the full numerical calculations. The power spectrum of temperature fluctuations due to the small scale ISW effect has wave number dependence $k^{-5}$ times that of the matter power spectrum.

## I. INTRODUCTION

It is well known that several effects contribute to Cosmic Microwave Background (CMB) fluctuations, *e.g.*, the difference in the gravitational potential between the last scattering surface (LSS) and the present (the Sachs-Wolfe effect) [1], Doppler shifts from moving electrons on the LSS [2], and the fluctuations corresponding to adiabatic or isocurvature initial conditions. The gravitational redshift induced by a time varying potential, *i.e.*, the *integrated Sachs-Wolfe* (ISW) effect, gives another important contribution to CMB fluctuations although it vanishes in linear theory for an $\Omega_0 = 1$ flat universe. This effect is dominant on very large scales [3], but only the non-linear effect is usually considered important on small scales [4]. In this paper, however, it is shown that the ISW effect does indeed lead to significant *small scale* CMB fluctuations for curvature or cosmological constant dominated models. In particular, under the assumption of early reionization, this effect is quite important since the Doppler contributions are severely cancelled on scales below the thickness of the LSS [5]. Such early reionization is actually required for open primeval isocurvature baryon (PIB) models [6]. We introduce here a simple model-independent analytic method for calculating this effect which agrees quite well with the detailed numerical solution.

## II. ANALYTIC FORMALISM

The full *first order* Boltzmann equation [7] for the evolution of temperature perturbations $\Theta(\mathbf{x}, \boldsymbol{\gamma}, \eta)$ in the gauge invariant notation [8] is given by

$$\dot{\Theta} + \dot{\Psi} + \gamma^i \frac{\partial}{\partial x^i}(\Theta + \Psi) - \Gamma^i_{jk}\gamma^j\gamma^k \frac{\partial}{\partial \gamma^i}(\Theta + \Psi) = \dot{\Psi} - \dot{\Phi} + \dot{\tau}(\Theta_0 - \Theta + \gamma^i v_i + \frac{1}{16}\gamma^i\gamma^j \Pi_{ij}), \qquad (1)$$

where $v_i$ is the matter velocity ($c=1$), $\Pi_{ij}$ is the anisotropic stress perturbation, $\gamma_i$ are the direction cosines for the photon momentum, $\Gamma$ is the Christoffel symbol, over-dots are derivatives with respect to conformal time $\eta = \int dt/a$, $\Theta_0$ is the monopole component of $\Theta$, and $\Psi$ is the Newtonian potential. The last term in eqn. (1) accounts for Compton scattering, where $\dot{\tau} = x_e n_e \sigma_T a$ is the differential optical depth, with $x_e$ the ionization fraction, $n_e$ the electron number density, and $\sigma_T$ the Thomson cross section. The perturbation to the intrinsic spatial curvature $\Phi$ is given by the Poisson equation $(\nabla^2 - 3K)\Phi = 4\pi G a^2 \rho \delta$, where the curvature constant $K = -H_0^2(1 - \Omega_0 - \lambda)$, the Hubble constant $H_0 = 100h$ km s$^{-1}$ Mpc$^{-1}$, the scaled cosmological constant $\lambda = \Lambda/3H_0^2$, and the total density fluctuation $\delta = \delta\rho/\rho$.

In the absence of scattering, the right hand side of (1) vanishes for a flat ($\Omega_0 = 1$) universe since $\dot{\Phi} = -\dot{\Psi} = 0$ in first order. We then obtain the usual free streaming equations and the ordinary Sachs-Wolfe effect, *i.e.* $\Theta(\mathbf{x}, \boldsymbol{\gamma}, \eta_0) + \Psi[\mathbf{x}(\eta_0), \eta_0] = \Theta[\mathbf{x}(\eta_*), \boldsymbol{\gamma}, \eta_*] + \Psi[\mathbf{x}(\eta_*), \eta_*]$, where $\eta_0$ and $\eta_*$ are the present and the last scattering epoch respectively. The more conventional expression $^1/_3\Psi(k, \eta_*)$ is obtained from a combination of $\Theta$ and $\Psi$ on the LSS (see [7]). Moreover, the quantity $\Theta + \Psi$ is the temperature anisotropy apart from the unobservable present day monopole and dipole. In open and (flat) lambda dominated universes, $\dot{\Psi} - \dot{\Phi}$ does not vanish and an additional effect, the ISW effect arises.

It is possible to obtain a simple approximate analytic solution to eqn. (1) by analogy to thick-LSS techniques [9, 10]. In a pressureless matter dominated universe, $\Phi = -\Psi$ [7], and on small scales, we may drop the curvature terms. We can then solve this equation in Fourier space after Compton drag has become negligible $z < z_d \simeq 130(\Omega_0 h^2)^{1/5} x_e^{-2/5}$:

$$\Theta(k, \mu, \eta_0) + \Psi(k, \eta_0) \simeq [\Theta(k, \mu, \eta_d) + \Psi(k, \eta_d)]e^{ik\mu(\eta_d - \eta_0)}e^{-\tau(\eta_i, \eta_0)} + \Theta_{SSF} + \Theta_{ISW}, \qquad (2)$$

$\tau(\eta_1, \eta_2) = \int_{\eta_1}^{\eta_2} \dot{\tau}(\eta)d\eta$ is the optical depth, $k\mu = \mathbf{k} \cdot \boldsymbol{\gamma}$, $\Theta_{SSF}$ is the secondary scattering fluctuation (SSF) generated on the new LSS if there is reionization, and $\Theta_{ISW}$ is the ISW effect. Explicitly, these terms are given by

$$\Theta_{SSF} = \int_{\eta_d}^{\eta_0} (\Theta_0 + \Psi + \mu v)\dot{\tau}e^{-\tau(\eta, \eta_0)}e^{ik\mu(\eta - \eta_0)}d\eta,$$
$$\Theta_{ISW} = 2\int_{\eta_d}^{\eta_0} \dot{\Psi}(k, \eta)e^{-\tau(\eta, \eta_0)}e^{ik\mu(\eta - \eta_0)}d\eta, \qquad (3)$$

where $\Theta_{SSF}$ has contributions from the Doppler [9] and ordinary Sachs-Wolfe effect.

Note that eqns. (2), (3), and the following argument are valid for *any* ionization history. In particular, for standard recombination $x_e(z < z_d \simeq 1000) = 0$, and $\tau(\eta_d, \eta_0) = \tau(\eta, \eta_0) = \Theta_{SSF} = 0$, *i.e. only* the ISW effect adds to $\Theta(k, \mu, \eta_d)$, the primary temperature fluctuations after standard last scattering. On the other hand, for reionized scenarios, the primary fluctuations are damped by $\exp(-\tau)$ due to rescattering [11]; in no-recombination scenarios, they may be ignored since $\tau(\eta_d, \eta_0) \gg 1$.

Whereas the integrand of $\Theta_{SSF}$ peaks at the LSS, the integrand of $\Theta_{ISW}$ is localized near the curvature dominated epoch $(1 + z_g) \simeq 1/\Omega_0 - 1$, or lambda dominated epoch $(1 + z_g) \simeq [1/\Omega_0 - 1]^{1/3}$. This epoch defines what may be called a *gravitational "last scattering" surface* (GLSS). A simple and useful analytic solution for $\Theta_{ISW}$ can be obtained on scales smaller than the thickness of the GLSS by noting that the LSS cancellation arguments for the SSF terms (valid on scales smaller than the thickness of the LSS) [5,11] can apply to the GLSS as well.

Extending the techniques of Kaiser [9] and Efstathiou [10] to the ISW effect, we find that the present day power spectrum of temperature fluctuations

$$k^3 \mathcal{P}(k) \equiv \frac{k^3}{2\pi^2} |\Theta + \Psi|^2_{rms} \simeq \frac{2}{\pi} \frac{V_x}{\eta_0^3} \frac{P(k)}{(k\eta_0)^2} \int_0^1 |G_{SSF}(x) + G_{ISW}(x)|^2 dx, \tag{4}$$

where $|\Theta+\Psi|^2_{rms} = \frac{1}{2} \int_{-1}^1 d\mu |\Theta(k,\mu,\eta_0) + \Psi(k,\eta_0)|^2$, $x = \eta/\eta_0$, the matter power spectrum $P(k) \equiv \langle|\delta(k)|^2\rangle$,

$$\begin{aligned} G_{SSF}(x) &= \frac{1}{2}\eta_0^3(\ddot{D}\dot{\tau} + \dot{D}\ddot{\tau})e^{-\tau(\eta,\eta_0)}, \\ G_{ISW}(x) &= \frac{3}{2}a^{-2}\eta_0^3 H_0^2 \Omega_0 (\dot{D}a - D\dot{a})e^{-\tau(\eta,\eta_0)}, \end{aligned} \tag{5}$$

$a = 1/(1+z)$, and $D$ is the growth factor in linear theory [12] such that $\delta(\eta) = D(\eta)\delta(\eta_0)$.

Notice that the wave number dependence of the ISW contribution is identical to the secondary Doppler term: $\mathcal{P}(k) \propto k^{-5} P(k)$. This dependence is easy to understand physically. Only the $\mu = 0$ mode in $\Theta$ can survive cancellation across a thick LSS or GLSS since the contributions are then additive. In general this restriction suppresses fluctuations by a factor $(k\eta_0)^{-1/2}$. However, for the Doppler and ordinary Sachs-Wolfe contributions, there is an additional suppression by a factor $(k\eta_0)^{-1}$ since both sources in the original equation (1) are proportional to $\mu$. Generalizing refs. [9, 10], we have set $\dot{\tau}(\Theta_0 + \Psi + \mu v) = ik^{-1}[\dot{v}\dot{\tau} + v\ddot{\tau} + \mathcal{O}(\dot{\tau}^2 \Psi) + \mathcal{O}(\dot{\tau}\dot{\Psi})]$. Since $\Psi \propto v/k$, the ordinary Sachs-Wolfe term is negligible on small scales compared with Doppler term. On the other hand, the integrated Sachs-Wolfe term, which does not vanish for the $\mu = 0$ mode, has the *same* dependence in $k$ as the "cancelled" Doppler term. It is therefore not negligible on small scales.

### III. COMPARISON WITH NUMERICAL RESULTS

For comparison, we solve eqn. (1) numerically up to present by the method outlined in [13]. It is instructive to examine first the time evolution of the quantities in eqn. (1). In Figure 1, we have chosen a representative no-recombination model (see caption) to illustrate the evolution of $\Psi$, $v$, and $|\Theta + \Psi|_{rms}$ in (a) an open and (b) a lambda ($\lambda = 1 - \Omega_0$) model. After the drag epoch $z \lesssim 100$, $v$ and $\Psi$ evolve in linear theory. At the curvature or lambda dominated epoch, $\Psi$ starts to decay, marking out the GLSS. More precisely, whereas the LSS is defined by the visibility function $\dot{\tau}\exp(-\tau)$, the GLSS is given by $\dot{\Psi}\exp(-\tau)$ (dashed lines, arbitrary normalization).

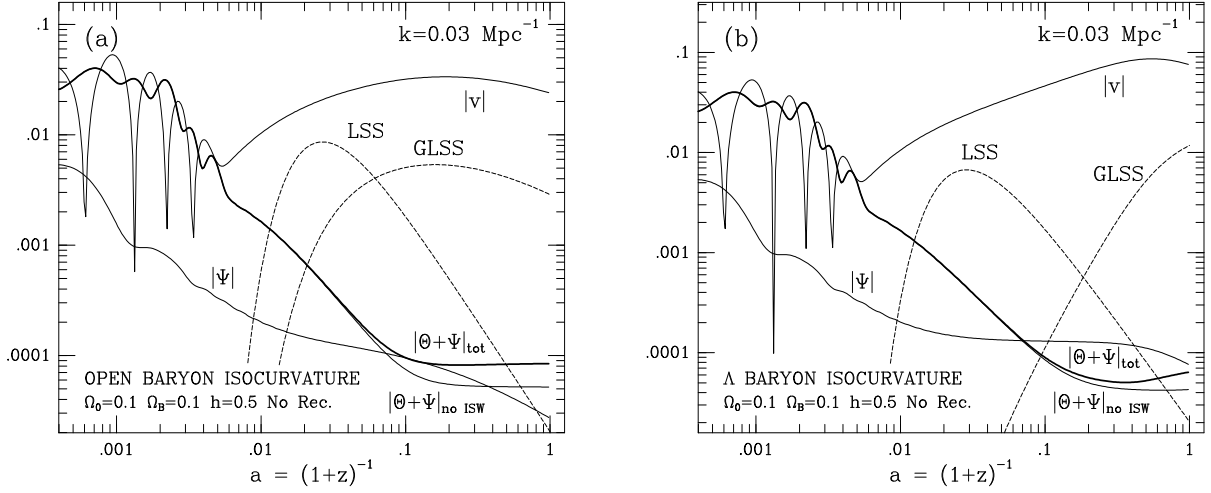

FIG 1: Evolution of $\Psi, v$, and the r.m.s. of $\Theta + \Psi$ with and without ISW from numerical calculations for a specific small scale mode ($k = 0.03\text{Mpc}^{-1}$) in (a) an open model ($\lambda = 0$) and (b) a lambda model ($\lambda = 1 - \Omega_0$) for an $\Omega_0 = \Omega_B = 0.1$, $h = 0.5$ universe with no recombination and isocurvature initial conditions. Overall normalization is arbitrary. Through the LSS, temperature fluctuations, originally of $\mathcal{O}(v)$, become increasingly suppressed due to cancellation and rescattering. We have also plotted the shape of the LSS [$\propto \dot\tau \exp(-\tau)$] and the GLSS [$\propto \dot\Psi \exp(-\tau)$] from linear theory for reference (dotted lines). In lambda models, the ISW effect contributes strongly near the present epoch, and the GLSS is not well defined. Nevertheless the analytic approximations still work quite well.

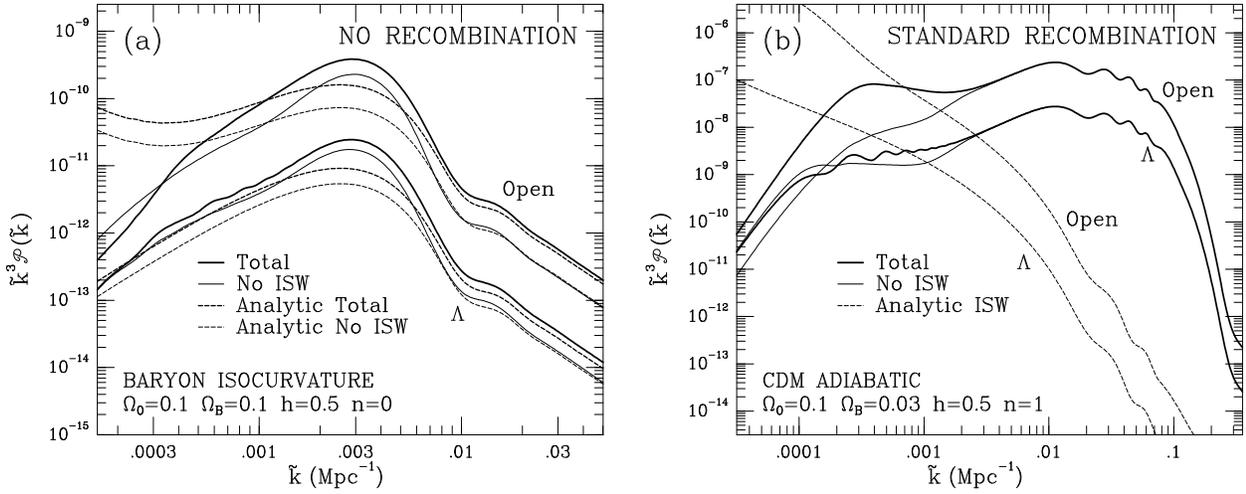

FIG 2: Power spectra for the temperature fluctuations $\tilde{k}^3 \mathcal{P}(\tilde{k})$ where $\tilde{k}^2 = k^2 + K$ [3]. from analytic and numerical calculations: (a) No-recombination; open (upper set) and lambda (lower set) baryon isocurvature models with spectral index $n = 0$ and parameters $\Omega_0 = \Omega_B = 0.1$, $h = 0.5$ where the initial entropy fluctuations $|S(\tilde{k})|^2 \propto \tilde{k}^n$. The solid lines represent the numerical results with (thick) and without (thin) ISW. The dashed lines represent analytic calculations in an analogous manner. (b) Standard recombination; open (upper set) and lambda (lower set) CDM models with adiabatic initial conditions $P(\tilde{k}) \propto \tilde{k}^n$ and parameters $\Omega_0 = 0.1, \Omega_B = 0.03, h = 0.5, n = 1$. On large scales, the power spectra fall off since we have subtracted the unobservable monopole and dipole.

We have plotted the resultant temperature fluctuation $|\Theta + \Psi|_{rms}$ with (thick line) and without (thin line) the ISW term in eqn. (1). Notice that the ISW has a negligible effect during the last scattering epoch; only in the lambda or curvature dominated epoch *and* after suppression through the LSS does the ISW effect become important. In these no-recombination scenarios, the LSS and GLSS have comparable width, and the ISW effect is important across the spectrum.

We plot the full present day power spectrum of temperature fluctuations with a $\sigma_8 = 1$ normalization for various models in Figure 2. In panel (a), the results for no-recombination isocurvature open (top lines) and lambda models (bottom lines) of Figure 1 are shown. The analytic calculations (dashed line) with (thick) and without (thin) ISW agree quite well with the full numerical solution on small scales. Deviations on large scales are expected since the analytic formalism is only valid below the thickness of the LSS and GLSS. We calculate the ratios of the power spectrum of temperature fluctuations with and without the ISW contribution for other baryonic no-recombination models in Table 1. Notice that the ISW effect gives a contribution of the same order as the secondary Doppler effect for low $\Omega_0$ models.

**TABLE I**

| $\Omega_0$ | $\lambda$ | $R_{num}$ | $R_{anal}$ |
|---|---|---|---|
| 0.1 | 0.0 | 2.6 | 2.2 |
| 0.2 | 0.0 | 2.1 | 1.9 |
| 0.4 | 0.0 | 1.6 | 1.5 |
| 0.6 | 0.0 | 1.3 | 1.3 |
| 0.8 | 0.0 | 1.2 | 1.1 |
| 1.0 | 0.0 | 1.0 | 1.0 |
| 0.1 | 0.9 | 1.9 | 1.7 |
| 0.2 | 0.8 | 1.5 | 1.4 |
| 0.4 | 0.6 | 1.2 | 1.1 |
| 0.6 | 0.4 | 1.1 | 1.1 |
| 0.8 | 0.2 | 1.1 | 1.0 |

TAB 1: Ratios of the power in temperature fluctuations with to without ISW $R \equiv \mathcal{P}_{tot}/\mathcal{P}_{noISW}$ given by the numerical ($R_{num}$) and analytical ($R_{anal}$) computations for an $\Omega_B = \Omega_0$, $h = 0.5$ universe. $R_{num}$ is calculated for $k = 1\Omega_0 h^2 \text{Mpc}^{-1}$ but does not vary significantly on small scales. The ISW effect is less significant in lambda models since lambda domination occurs at a relatively late epoch. Analytic predictions agree quite well with numerical results ($\lesssim 20\%$ in power, $\lesssim 10\%$ in temperature fluctuations).

In Fig. 2b, we examine the ISW effect in an open (top lines) and lambda model (bottom lines) with standard recombination. Here we take an adiabatic CDM model for which this ionization history is more natural. Although this model overpredicts fluctuations if normalized to $\sigma_8$, we choose it to demonstrate that even in this extreme case, the ISW effect plays no role on small scales. For standard recombination, the LSS is relatively thin, and the ISW effect is masked until the Silk damping scale. Furthermore since the GLSS is much thicker than the LSS, cancellation of the ISW effect begins at a much larger scale and therefore is negligible at the Silk damping scale. Note however that the ISW effect *itself* does not depend sensitively on the ionization history. Moreover, even in these models, the ISW plays an important role on large scales (where the analytic formalism breaks down). It is interesting that the analytic formalism nevertheless predicts the correct scale for dominance of the ISW effect. Finally, we should mention that the ISW *does* contribute significantly to open or lambda adiabatic CDM models with sufficiently early reionization or no recombination.

## IV. DISCUSSION

We have found that the ISW effect is important at *all* scales in a reionized open or lambda universe including the so-called Doppler peak and below. Furthermore, we have developed an analytic formalism which compares quite well with detailed numerical solutions. From this formalism, we find that the power spectrum of temperature fluctuations behaves on small scales as $\mathcal{P}(k) \propto k^{-5} P(k)$ for *both* the Doppler and ISW contributions, despite the fact that the scale dependence of $v$ and $\dot{\Psi}$ themselves are not similar, due to a difference in their cancellation behavior. The amplitude of the temperature fluctuations can be predicted to 10% accuracy by the analytic approximation. In standard recombination models, the small scale ISW effect is masked since the LSS is relatively thin compared to the GLSS, and thus cancellation of the Doppler effect does not occur until a very small scale. The analytic approach derived in this *Letter* is a potentially powerful tool for understanding and calculating the ISW effect. It is independent of the detailed model and is thus applicable to all scenarios which have a time varying potential, *e.g.*, even topological defect models [14] for which the ISW effect is essentially the only important contribution to CMB fluctuations.

## ACKNOWLEDGEMENTS

We would like to thank F. Atrio-Barandela, D. Scott and J. Silk for useful comments and discussions. W.H. has been partially supported by an NSF fellowship. N.S. acknowledges financial support from a JSPS postdoctoral fellowship for research abroad.

**TABLE I**

| $\Omega_0$ | $\lambda$ | $R_{num}$ | $R_{anal}$ |
|---|---|---|---|
| 0.1 | 0.0 | 2.6 | 2.2 |
| 0.2 | 0.0 | 2.1 | 1.9 |
| 0.4 | 0.0 | 1.6 | 1.5 |
| 0.6 | 0.0 | 1.3 | 1.3 |
| 0.8 | 0.0 | 1.2 | 1.1 |
| 1.0 | 0.0 | 1.0 | 1.0 |
| 0.1 | 0.9 | 1.9 | 1.7 |
| 0.2 | 0.8 | 1.5 | 1.4 |
| 0.4 | 0.6 | 1.2 | 1.1 |
| 0.6 | 0.4 | 1.1 | 1.1 |
| 0.8 | 0.2 | 1.1 | 1.0 |

**TABLE CAPTIONS**

TAB 1: Ratios of the power in temperature fluctuations with to without ISW $R \equiv \mathcal{P}_{tot}/\mathcal{P}_{noISW}$ given by the numerical ($R_{num}$) and analytical ($R_{anal}$) computations for an $\Omega_B = \Omega_0$, $h = 0.5$ universe. $R_{num}$ is calculated for $k = 1\Omega_0 h^2 \mathrm{Mpc}^{-1}$ but does not vary significantly on small scales. The ISW effect is less significant in lambda models since lambda domination occurs at a relatively late epoch. Analytic predictions agree quite well with numerical results ($\lesssim 20\%$ in power, $\lesssim 10\%$ in temperature fluctuations).

**FIGURES:**

FIG 1: Evolution of $\Psi, v$, and the r.m.s. of $\Theta + \Psi$ with and without ISW from numerical calculations for a specific small scale mode ($k = 0.03 \mathrm{Mpc}^{-1}$) in (a) an open model ($\lambda = 0$) and (b) a lambda model ($\lambda = 1 - \Omega_0$) for an $\Omega_0 = \Omega_B = 0.1$, $h = 0.5$ universe with no recombination and isocurvature initial conditions. Overall normalization is arbitrary. Through the LSS, temperature fluctuations, originally of $\mathcal{O}(v)$, become increasingly suppressed due to cancellation and rescattering. We have also plotted the shape of the LSS [$\propto \dot{\tau} \exp(-\tau)$] and the GLSS [$\propto \dot{\Psi} \exp(-\tau)$] from linear theory for reference (dotted lines). In lambda models, the ISW effect contributes strongly near the present epoch, and the GLSS is not well defined. Nevertheless the analytic approximations still work quite well.

FIG 2: Power spectra for the temperature fluctuations $\tilde{k}^3 \mathcal{P}(\tilde{k})$ where $\tilde{k}^2 = k^2 + K$ [3]. from analytic and numerical calculations: (a) No-recombination; open (upper set) and lambda (lower set) baryon isocurvature models with spectral index $n = 0$ and parameters $\Omega_0 = \Omega_B = 0.1$, $h = 0.5$ where the initial entropy fluctuations $|S(\tilde{k})|^2 \propto \tilde{k}^n$. The solid lines represent the numerical results with (thick) and without (thin) ISW. The dashed lines represent analytic calculations in an analogous manner. (b) Standard recombination; open (upper set) and lambda (lower set) CDM models with adiabatic initial conditions $P(\tilde{k}) \propto \tilde{k}^n$ and parameters $\Omega_0 = 0.1, \Omega_B = 0.03, h = 0.5, n = 1$. On large scales, the power spectra fall off since we have subtracted the unobservable monopole and dipole.